\documentclass[pra, 10pt, twocolumn, letter, floatfix]{revtex4}
\usepackage{amsmath}
\usepackage{graphicx}
\usepackage{rotating}
\usepackage{bm}
\allowdisplaybreaks
\usepackage{setspace}

\begin{document}



\vphantom{Hello}

\vphantom{Hello}

\vphantom{Hello}

\vphantom{Hello}

\vphantom{Hello}

\title{Optimizing large parameter sets in variational quantum Monte Carlo}
\author{Eric Neuscamman$\mathrm{{}^\ast}$, C. J. Umrigar$\mathrm{{}^\dag}$, and Garnet Kin-Lic Chan$\mathrm{{}^\ddag}$}
\affiliation{$\mathrm{{}^\ast}$Department of Chemistry, University of California, Berkeley, California 94720\\
             $\mathrm{{}^\dag}$Department of Physics, Cornell University, Ithaca, New York 14853\\
             $\mathrm{{}^\ddag}$Department of Chemistry and Chemical Biology, Cornell University, Ithaca, New York 14853}
\date{\today}

\begin{abstract}
We present a technique for optimizing hundreds of thousands of variational parameters in variational quantum Monte Carlo.
By introducing iterative Krylov subspace solvers and by multiplying by the Hamiltonian and overlap matrices as they are
sampled, we remove the need to construct and store these matrices and thus bypass the most expensive steps of the stochastic
reconfiguration and linear method optimization techniques.
We demonstrate the effectiveness of this approach by using stochastic reconfiguration to optimize a correlator product state
wavefunction with a pfaffian reference for four example systems.
In two examples on the two dimensional Hubbard model, we study 16 and 64 site lattices, recovering energies accurate to
1\% in the smaller lattice and predicting particle-hole phase separation in the larger.
In two examples involving an ab initio Hamiltonian, we investigate the potential energy curve of a symmetrically
dissociated 4x4 hydrogen lattice as well as the singlet-triplet gap in free base porphin.
In the hydrogen system we recover 98\% or more of the correlation energy at all geometries, while for porphin we
compute the gap in a 24 orbital active space to within 0.02eV of the exact result.
The numbers of variational parameters in these examples range from $4\times10^3$ to $5\times10^5$, demonstrating
an ability to go far beyond the reach of previous formulations of stochastic reconfiguration.

\vphantom{Hello}
\end{abstract}

\maketitle

\section{Introduction}
\label{sec:introduction}

Quantum Monte Carlo (QMC) is a powerful technique for extracting predictions from the electronic Schr\"odinger
equation \cite{FouMitNeeRaj-RMP-01}.
The variational (VMC) and diffusion (DMC) Monte Carlo methods in particular can produce highly accurate predictions
provided that a sufficiently flexible trial wavefunction is available and that the variational parameters of this
wavefunction can be optimized.
However, VMC and DMC suffer from the major limitation that the most effective stochastic optimization algorithms cannot
handle more than a few thousand variational parameters.
These algorithms, which include the Newton \cite{UmrFil-PRL-05}, approximate Newton \cite{Sor-PRB-05},
linear (LM) \cite{Nightingale:2001:linear_method,TouUmr-JCP-07,TouUmr-JCP-08,UmrTouFilSorHen-PRL-07}
and stochastic reconfiguration (SR) \cite{Sorella:2001:SR} methods, are currently constrained by their need to build and
store matrices that become unmanageable when the number of variational parameters becomes large.
Other stochastic optimization algorithms \cite{Sandvik:2007:stoch_opt} that rely only on stochastic estimates for the
energy gradient can treat more variational parameters, but their steepest-descent character makes for less efficient
convergence to the energy minimum, especially compared to the LM.
In order to make effective use of sophisticated trial wavefunctions such as tensor networks, which can contain millions
of variational parameters, it is imperative that more capable optimization methods be developed.

The LM and SR optimization methods reduce to solving either a system of linear equations or a linear
eigenvalue problem in which the matrices in question are determined by stochastic sampling.
The essential difficulty in this approach is that the dimension of these matrices is equal to the number of
variational parameters, preventing their construction when there are more than a few thousand variables.
Here we propose solving the optimization methods' central linear problems using iterative Krylov subspace
algorithms, which do not require the matrices to be built explicitly.
Instead, these solvers require that one evaluate matrix-vector products, which we will show to be far less
difficult than actually building the relevant matrices.
In VMC this approach is made particularly efficient by the strategy of operating by the matrices during the
sampling process, as each sampled configuration contributes an outer product to the overall matrix, and
outer products are particularly easy to operate by.

In this paper we will demonstrate this approach by using the conjugate gradient (CG) iterative solver to improve the
SR method.  We also derive a method for improving the LM using the generalized Davidson solver,
although we will only present numerical results for SR (a computer implementation for the LM is underway).
We will begin by developing the theory for the accelerated SR and LM and also for the particular
wavefunction ansatz that we employ.
After developing the theory, we will present numeric results for the SR method in four examples:
(a) the Hubbard model on a 4x4 lattice, (b) phase separation behavior in the 8x8 Hubbard model, (c) the potential
energy curve of a symmetrically dissociated 4x4 hydrogen lattice, and (d) the singlet-triplet gap of free base
porphin.
Note that the numerical studies carried out here are primarily concerned with the optimization problem.
A detailed examination of the physics of these examples will be carried out elsewhere.

\clearpage

\section{Theory}
\label{sec:theory}

\subsection{Accelerated Stochastic Reconfiguration}
\label{sec:accelerated_sr}

The SR method can be viewed as an approximate imaginary time evolution in a specially
chosen subspace $\Omega$ of the full Hilbert space.
For a wavefunction $|\Psi(\alpha_1, \alpha_2, \ldots)\rangle$ with variational parameters $\bm{\alpha}$,
this subspace is spanned by the wavefunction and its $\bm{\alpha}$-derivatives,
\begin{align}
\label{eqn:sr_subspace}
\Omega = \mathrm{span}\left(~|\Psi^0\rangle,~|\Psi^1\rangle,~|\Psi^2\rangle,~\ldots~\right),
\end{align}
where $|\Psi^0\rangle\equiv|\Psi\rangle$ and $|\Psi^i\rangle\equiv\partial|\Psi\rangle/\partial\alpha_i$
for $i>0$.
The strategy of the SR method is to minimize the wavefunction's energy by repeatedly operating by
$T=1-\tau H$ (the imaginary time evolution operator $e^{-\tau H}$ expanded to first order), where $\tau$
is a small number and $H$ is the Hamiltonian.
After each application of $T$, the result is projected into $\Omega$ to produce a new wavefunction of the
form $|\Psi^\prime\rangle=\sum_i x_i |\Psi^i\rangle$, in which the coefficients $\bm{x}$ are given by
\begin{align}
\label{eqn:sr_lin_eq}
\langle\Psi^i|\left(1-\tau H\right)|\Psi\rangle = \sum_j \langle\Psi^i|\Psi^j\rangle x_j.
\end{align}
Finally, because $\tau$ is small, the new wavefunction $|\Psi^\prime\rangle$ can be closely approximated by
$|\Psi(\alpha_1^\prime, \alpha_2^\prime, \ldots)\rangle$, where $\alpha_i^\prime = \alpha_i + x_i / x_0$.
To summarize, one solves the linear equation given in Eq.\ (\ref{eqn:sr_lin_eq}) and updates $\bm{\alpha}$
accordingly, after which the subspace $\Omega$ is redefined for the new wavefunction.
This entire procedure is repeated until the energy of the wavefunction has converged.

Previously, the SR overlap matrix $S_{i j} = \langle\Psi^i|\Psi^j\rangle$ was constructed explicitly.
Here we will avoid building $\bm{S}$ entirely, relying instead on the CG algorithm to solve Eq.\ (\ref{eqn:sr_lin_eq}).
This method proceeds iteratively, using information gained from a series of matrix-vector multiplications to
successively refine an approximation to the solution $\bm{x}$ in a space of orthonormal conjugate vectors.
The iteration proceeds until an arbitrary accuracy is achieved and typically converges in a number of steps far
smaller than the dimension of the matrix.
To see the advantages of using CG, consider the following expressions showing how the overlap matrix was previously
constructed through stochastic sampling,
\begin{align}
\label{eqn:sr_overlap}
\frac{S_{i j}}{\langle\Psi|\Psi\rangle} & = \sum_{\bm{n}} \frac{|\Psi_{\bm{n}}|^2}{\langle\Psi|\Psi\rangle}
                                                          \left( \frac{\Psi^i_{\bm{n}}}{\Psi_{\bm{n}}} \right)
                                                          \left( \frac{\Psi^j_{\bm{n}}}{\Psi_{\bm{n}}} \right), \\
|\Psi\rangle & = \sum_{\bm{n}} \Psi_{\bm{n}}|\bm{n}\rangle, \\
|\Psi^i\rangle & = \sum_{\bm{n}} \Psi^i_{\bm{n}}|\bm{n}\rangle.
\end{align}
Here a resolution of the identity $\sum_{\bm{n}} |\bm{n}\rangle\langle\bm{n}|$ has been inserted, creating a summation
over all possible system configurations $|\bm{n}\rangle$.
By multiplying and dividing by $|\Psi_{\bm{n}}|^2$ the summation has been formulated so that it can be evaluated
stochastically by sampling from the distribution $|\Psi_{\bm{n}}|^2/\langle\Psi|\Psi\rangle$.
However, building $\bm{S}$ stochastically using Eq.\ (\ref{eqn:sr_overlap}) takes at least $O(n_s n_v^2)$ time,
where $n_s$ is the number of samples and $n_v$ is the number of variational parameters.
Using the CG algorithm we may avoid this cost by instead evaluating matrix-vector products of the form $\bm{S}\bm{z}$.
As with the expression for constructing $\bm{S}$, this expression can be evaluated by stochastic sampling if we
insert a resolution of the identity,
\begin{align}
\label{eqn:sr_multiply_1}
\sum_j \frac{S_{i j}}{\langle\Psi|\Psi\rangle} z_j = \sum_j \sum_{\bm{n}} \frac{|\Psi_{\bm{n}}|^2}{\langle\Psi|\Psi\rangle}
                                                            \left( \frac{\Psi^i_{\bm{n}}}{\Psi_{\bm{n}}} \right)
                                                            \left( \frac{\Psi^j_{\bm{n}}}{\Psi_{\bm{n}}} \right) z_j.
\end{align}
By interchanging the order of summations we can rewrite this product as
\begin{align}
\label{eqn:sr_multiply_2}
\sum_j \frac{S_{i j}}{\langle\Psi|\Psi\rangle} z_j = \sum_{\bm{n}} \frac{|\Psi_{\bm{n}}|^2}{\langle\Psi|\Psi\rangle}
                                                            \frac{\Psi^i_{\bm{n}}}{\Psi_{\bm{n}}}
                                                            \left( \sum_j \frac{\Psi^j_{\bm{n}}}{\Psi_{\bm{n}}} z_j \right),
\end{align}
which can be evaluated in $O(n_s n_v)$ time provided that the derivative ratios $\Psi^i_{\bm{n}}/\Psi_{\bm{n}}$
have been pre-computed and stored, which is not difficult as the storage can be trivially divided between the
different processors.
While the CG algorithm does require multiple matrix-vector products to be evaluated, the number
of such products will be much smaller than $n_v$, greatly improving the efficiency of the SR method.

\begin{figure}[t]
\centering
\includegraphics[width=8.5cm,angle=0]{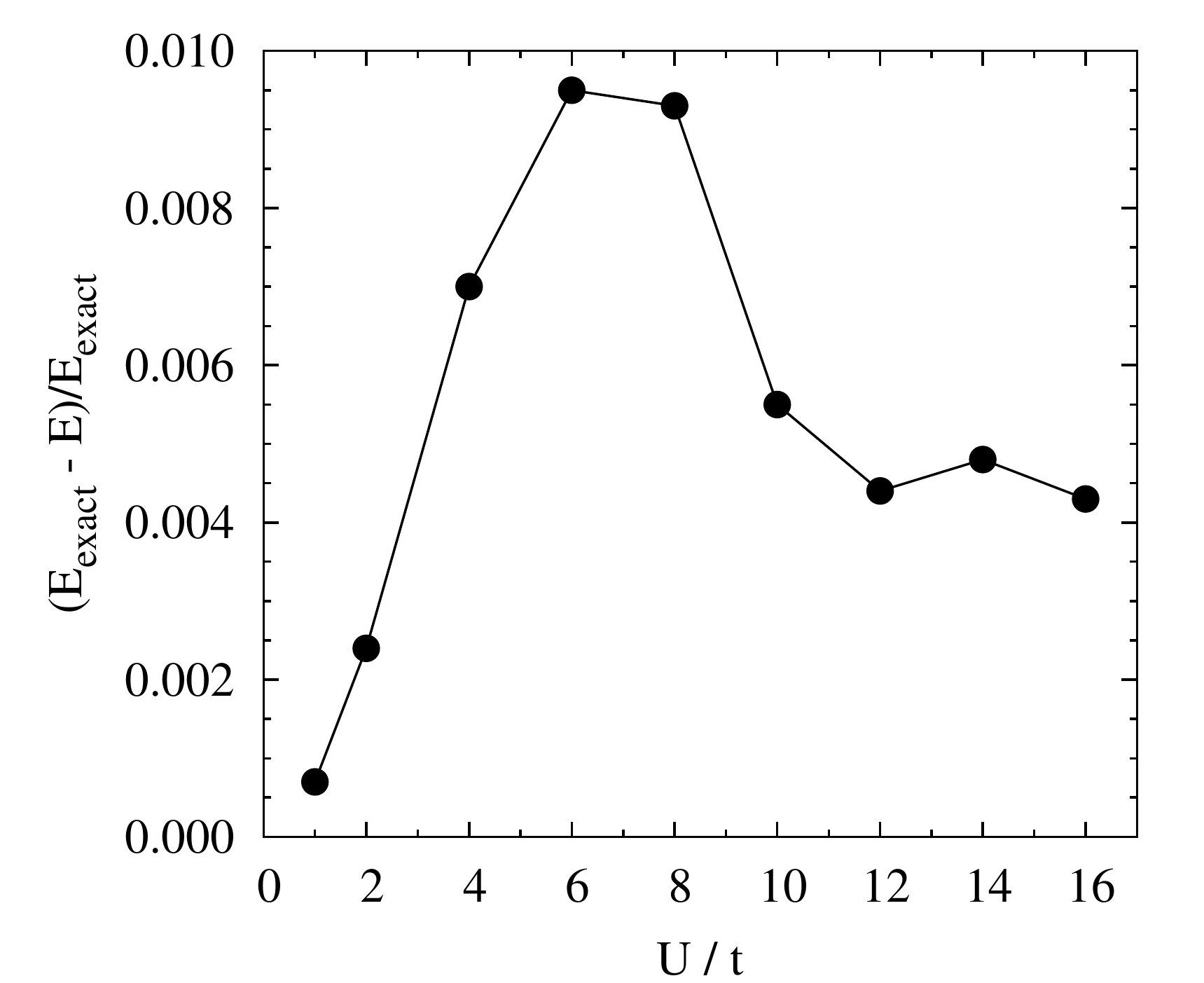}
\caption{Relative energy errors for the CPS-Pfaffian ansatz on a periodic 4x4 Hubbard lattice.
         Statistical errors are smaller than the symbol size and lines are guides to the eye.}
\label{fig:hubbard_4x4}
\end{figure}

\subsection{Accelerated Linear Method}
\label{sec:accelerated_linear}

The linear method (LM), formulated by Nightingale for linear parameters \cite{Nightingale:2001:linear_method}
and later extended to optimize nonlinear parameters \cite{TouUmr-JCP-07,TouUmr-JCP-08,UmrTouFilSorHen-PRL-07}, works in the
same subspace $\Omega$ as the SR method but typically converges more rapidly to the energy minimum.
It can be viewed as an approximate Newton method with a built in stabilization \cite{TouUmr-JCP-08} and often
converges even more rapidly than the Newton method.
Instead of using imaginary time evolution, the LM optimizes $|\Psi\rangle$ by finding the eigenstate of
lowest energy in the $\Omega$ subspace.
This eigenstate can be found by solving the following generalized eigenvalue problem,
\begin{align}
\label{eqn:linear_eigen}
\sum_j \langle\Psi^i| H |\Psi^j\rangle x_j = E \sum_k \langle\Psi^i|\Psi^k\rangle x_k,
\end{align}
where we now take $\bm{x}$ to be the coefficients of the desired eigenvector.
Once these coefficients are found, the variables $\bm{\alpha}$ can be updated to their new values
in the same manner as in SR, though care must be taken to check that for large parameter changes the resulting
parameters give an energy that is not higher outside of statistical errors.
If they do not, the step can be scaled down by using a line search, or, rotated and scaled down by adding a diagonal shift.
In practice, it is essential to modify the update in order to make it orthogonal to the original
wavefunction, a procedure that can be completed using the information resulting from a single
matrix-vector multiply involving the overlap matrix.

As with SR, the eigenvector $\bm{x}$ can be found without explicitly building the matrices $\bm{H}$
and $\bm{S}$ by using a Krylov subspace method, in this case the generalized Davidson algorithm
\cite{Morgan:1990:general_davidson}.
As with CG, it is sufficient to evaluate the matrix-vector products of $\bm{H}$ and $\bm{S}$ with arbitrary
trial vectors.
For $\bm{S}$, this product can be performed efficiently as explained above.
For $\bm{H}$, the difficulty of the multiplication depends on the complexity of the system's Hamiltonian,
but for the relatively general case of the non-relativistic Born-Oppenheimer Hamiltonian an efficient
evaluation is possible.
If we assume a fixed particle number, we may use a matrix factorization such as the Cholesky decomposition
\cite{Linderberg:1977:cholesky} to express this Hamiltonian as
\begin{align}
\label{eqn:cholesky_ham}
H = \sum_\mu \sum_{p q r s} L^\mu_{p q} R^\mu_{r s} a^\dag_p a_q a^\dag_r a_s,
\end{align}
where the operator $a^\dag_p$ ($a_p$) is the fermionic creation (destruction) operator for
the $p$th spin orbital, the index $\mu$ has a range of $O(n_o^2)$ ($n_o$ is the number of orbitals),
and the indices $p$, $q$, $r$, $s$ each have range $n_o$.
In practice, the range of $\mu$ can often be taken to be much smaller than $n_o^2$ while still
representing $H$ with sufficient accuracy.
By inserting an identity operator in the center of the Hamiltonian, the matrix-vector product on the left
hand side of Eq.\ (\ref{eqn:linear_eigen}) can be written as
\begin{align}
\label{eqn:linear_multiply}
& \frac{1}{\langle\Psi|\Psi\rangle} \sum_j \langle\Psi^i| H |\Psi^j\rangle x_j \\
\notag
& ~ = \sum_{j \bm{n} \mu p q r s} \frac{|\Psi_{\bm{n}}|^2}{\langle\Psi|\Psi\rangle}
                                     L^\mu_{p q} R^\mu_{r s}
                                     \frac{\langle\Psi^i|a^\dag_p a_q|\bm{n}\rangle}{\Psi_{\bm{n}}}
                                     \frac{\langle\bm{n}|a^\dag_r a_s|\Psi^j\rangle}{\Psi_{\bm{n}}}
                                     x_j \\
\notag
& ~ = \sum_{\bm{n}} \frac{|\Psi_{\bm{n}}|^2}{\langle\Psi|\Psi\rangle}
      \sum_{p q} Q_{\bm{n} q p i}
      \sum_\mu L^\mu_{p q}
      \sum_{r s} R^\mu_{r s}
      \sum_j Q_{\bm{n} r s j} x_j,
\end{align}
where we have defined the intermediate tensor
$Q_{\bm{n} r s j} = \langle\bm{n}|a^\dag_r a_s|\Psi^j\rangle / \Psi_{\bm{n}}$.
For the wavefunction presented in the next section, this intermediate can be
evaluated in $O(n_o^4)$ time for a given configuration $|\bm{n}\rangle$.
If we sample the configurations $|\bm{n}\rangle$ from the distribution
$|\Psi_{\bm{n}}|^2/\langle\Psi|\Psi\rangle$, we see that the entire matrix-vector
product can be evaluated in $O(n_s n_o^4)$ time by performing the summations in the
last line of Eq.\ (\ref{eqn:linear_multiply}) from right to left.
We therefore see that like SR, the LM can be performed without explicitly
constructing the matrices involved.

\begin{figure}[t]
\centering
\includegraphics[width=8.0cm,angle=0]{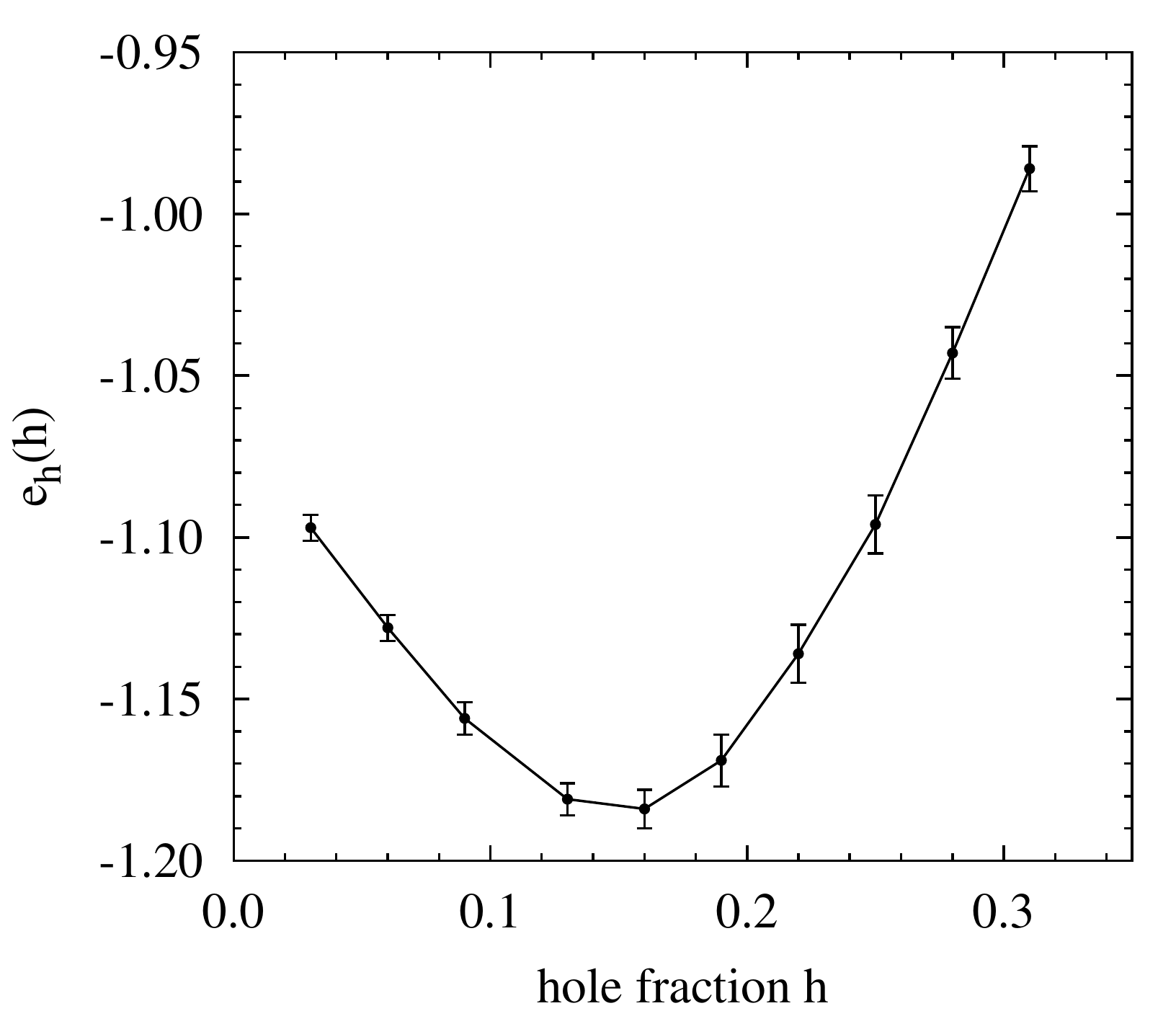}
\caption{The function $e_h(h)$ on an 8x8 Hubbard lattice with twist-averaged boundary conditions and $U/t=4$.
         The presence of a minimum implies that our ansatz predicts phase separation in the 2D Hubbard model.}
\label{fig:hubbard_8x8}
\end{figure}

\subsection{Wavefunction Ansatz}
\label{sec:ansatz}

For our variational ansatz, we use a product of a correlator product state (CPS) tensor network
\cite{CHANGLANI:2009:cps, MEZZACAPO:2009:entangled_plaquettes} and a pfaffian pairing wavefunction
\cite{Bajdich:2006:pfaffian,Bajdich:2008:pfaffian,Wimmer:2011:pfaffian}.
As discussed in Ref.\ \cite{Neuscamman:2010:pcps}, the CPS ansatz can be expressed as a product of correlators
acting on a reference wavefunction.
Here we take the same approach, but with a pfaffian as the reference rather than a Slater determinant.
The wavefunction is written as
\begin{align}
\label{eqn:ansatz}
|\Psi\rangle = \prod_p \hat{C}_p \left( \sum_{i < j} f_{i j} a^\dag_i a^\dag_j \right)^{N/2} |0\rangle,
\end{align}
where the operators $\hat{C}_p$ are correlators, $\bm{f}$ is the pairing matrix, $N$ is
the number of electrons, and $|0\rangle$ is the vacuum.
The indices $i,j$ range over all spin orbitals, so our pairing function creates both singlet
and triplet pairs, unlike the more restrictive antisymmetrized geminal power \cite{POPLE:1953:agp,COLEMAN:1965:AGP}.
Two typical types of correlators are long range pairs and $n \times n$ square plaquettes.
In each case, both the spin $\uparrow$ and $\downarrow$ versions of the spatial orbitals are included in
a correlator, so the number of contained spin orbitals (variational parameters) is 4 ($2^4$) for a pair
correlator and $2n^2$ ($2^{2n^2}$) for a plaquette.

\section{Results}
\label{sec:results}

Here we demonstrate the accelerated SR method by applying it to four example systems
in conjunction with our CPS-pfaffian ansatz.
In each system, we restrict our sampling to configurations with the correct total
number of electrons and the correct total $S_z$.
The accelerated LM has yet to be implemented on a computer and thus will be
tested in future work.

\subsection{4x4 Hubbard Model}
\label{sec:4x4_hubbard}

In our first example we studied a 4x4 Hubbard lattice at half filling with periodic boundary
conditions, which was chosen as it is an exactly soluble system that contains many of the
challenging features of the general 2D Hubbard model.
Two translationally invariant 3x3 correlators were used, one anchored on each sublattice,
giving a wavefunction with a total of 524,784 variational parameters.
In Figure \ref{fig:hubbard_4x4} we show the error relative to the exact result, which for
all ratios $U/t$ is less than 1\%.

\begin{figure}[b]
\centering
\includegraphics[width=8.5cm,angle=0]{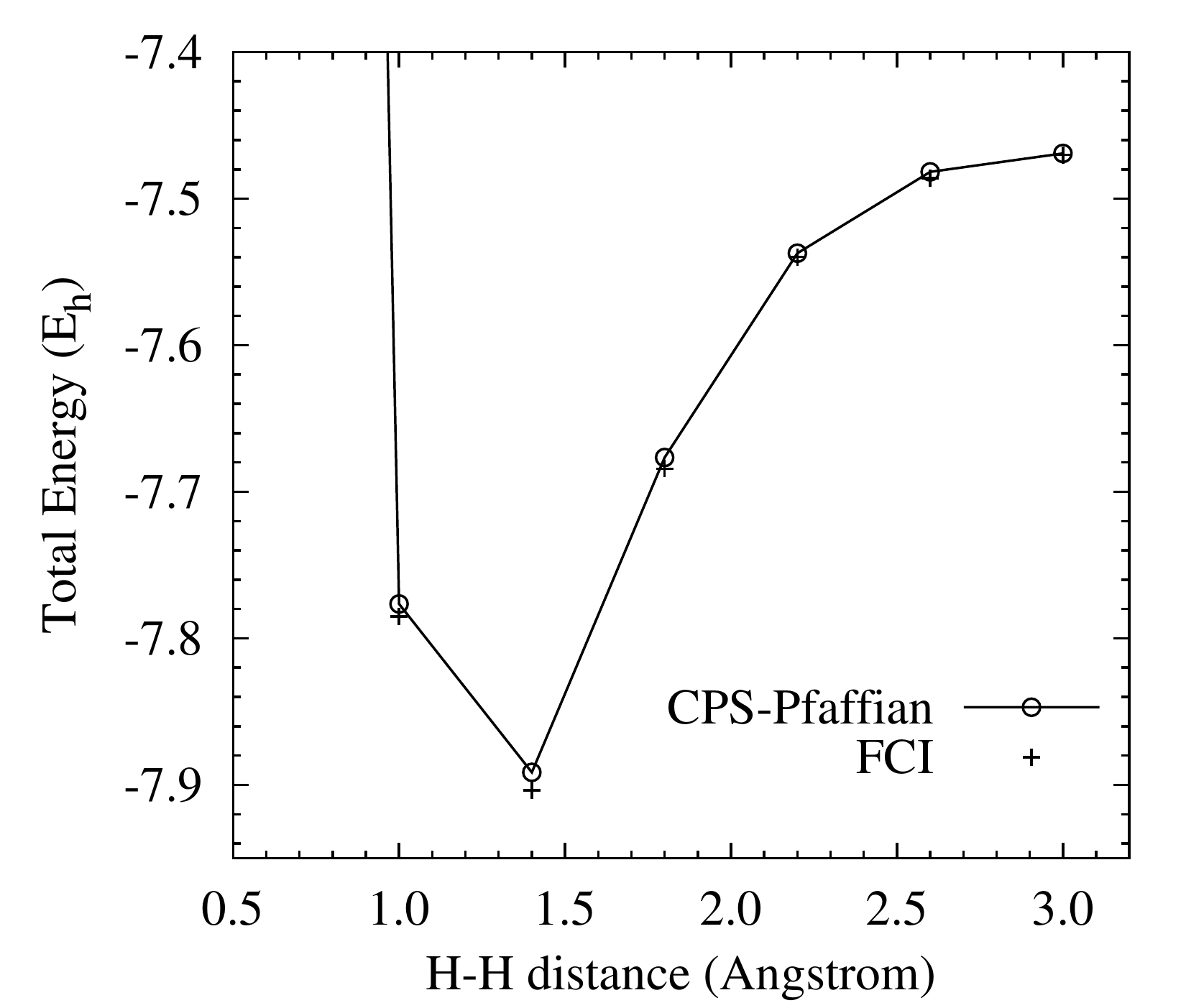}
\caption{Total energies of a 4x4 hydrogen lattice.
         Statistical errors are smaller than the symbol size and lines are guides to the eye.}
\label{fig:hydrogen_4x4}
\end{figure}
\begin{figure}[b]
\centering
\includegraphics[width=7.5cm,angle=0]{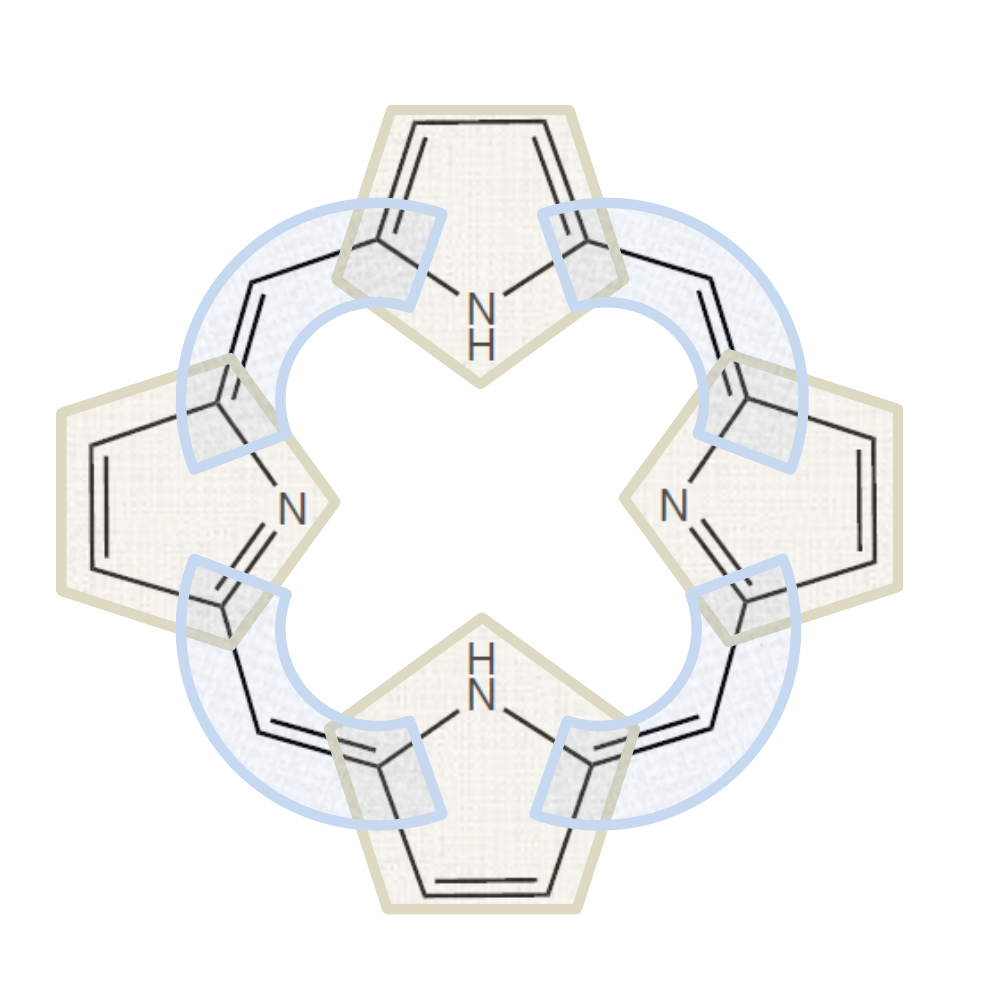}
\caption{In addition to all long range pairs, we use the correlators shown here when treating
         free base porphin.}
\label{fig:porphin}
\end{figure}

\subsection{8x8 Hubbard Model}
\label{sec:8x8_hubbard}

We have also applied our method to test for phase separation in the 2D Hubbard model, the exact
nature of which remains an interesting and unresolved problem in solid state physics.
To do so we studied an 8x8 lattice with twist-averaged boundary conditions (TABC)
\cite{Poilblanc:1991:tabc,Gros:1996:ibc,ZHANG:2008:hubbard_phase_sep} (we used 12 randomly chosen twists)
and $U/t=4$.
We used translationally invariant 2x2 and long range pair correlators, again using separate correlators
for each sublattice.
To check whether the system phase separates, we computed the quantity $e_h(h)$ employed in
Ref.\ \cite{ZHANG:2008:hubbard_phase_sep}, which will display a minimum at the critical hole density $h_c$
if phase separation occurs.
As seen in Figure \ref{fig:hubbard_8x8}, our approach predicts that the system will phase separate
with a critical hole density $0.14 < h_c < 0.15$.
This result provides a qualitative corroboration of the Constrained-Path Auxiliary Field QMC
\cite{ZHANG:2008:hubbard_phase_sep} results of Zhang et al, who predicted phase separation with
$h_c=0.1$ for the 8x8 lattice with TABC and $U/t=4$.

\subsection{4x4 Hydrogen Lattice}
\label{sec:4x4_hydrogen}

As an example of a strongly correlated problem involving an ab initio Hamiltonian, we have studied a 4x4
square lattice of hydrogen atoms in the STO-3G orbital basis \cite{Pople:1969:sto-3g} at various
nearest-neighbor distances.
As this system has open boundary conditions, we did not use translationally invariant correlators.
Instead, we used all 2x2 and long range pair correlators, which results in a wavefunction with 4,048 variational
parameters.
As seen in Figure \ref{fig:hydrogen_4x4}, the results closely match those of the exact wavefunction.
Even at the H-H distance with the worst error, our approach captures 98\% of the correlation energy, which we
define as the energy difference between the restricted Hartree Fock and exact wavefunctions.

\subsection{Free Base Porphin}
\label{sec:porphin}

As our final example, we computed the singlet-triplet gap of free base porphin in the 6-31G orbital basis
\cite{POPLE:1972:6-31g_basis}.
This system was chosen as an important quantum chemical problem for which exact results in the active space are
available for comparison.
For both the singlet and triplet wavefunctions, the 1s and $\sigma$ bonding orbitals resulting from a restricted
Hartree Fock calculation were treated as a closed shell determinant, while the 24 out-of-plane 2p orbitals from
the RHF solution were localized by the Pipek-Mezey \cite{PIPEK:1989:pipek_mezey_local} scheme to form an active
space containing the remaining 26 electrons.
This active space was treated with our CPS-pfaffian ansatz, with the correlators taken to be all pairs as well as
those shown in Figure \ref{fig:porphin}, for a total of 9,064 variational parameters.
Holding the core orbitals frozen, we computed an active space singlet-triplet gap of 1.77eV, which compares very
favorably with the converged spin-adapted density matrix renormalization group \cite{Sharma:20xx:sa-dmrg} result
of 1.75eV.

\section{Conclusions}
\label{sec:conclusions}

We have shown that by using the conjugate gradient iterative solver, it is possible to optimize hundreds of
thousands of variational parameters with the stochastic reconfiguration algorithm in the context of
variational Monte Carlo.
In addition, we have shown how the generalized Davidson solver can be used to provide a similar improvement
for the linear method.
Using our accelerated SR algorithm, we demonstrated that a CPS-pfaffian wavefunction ansatz is capable of
treating a number of challenging two dimensional systems that display both weakly and strongly correlated
physics.
Together, these advances provide a powerful new method for modeling both quantum chemical and solid state
systems.
In the future, we expect optimizations of millions of parameters to be possible, which will allow even more
sophisticated trial wavefunctions to be used in variational and diffusion Monte Carlo.

\section{Acknowledgments}
\label{sec:acknowledgments}

This work was supported by NSF grants CHE-1004603 and DMR-0908653 and by
the Miller Institute for Basic Research in Science.

\bibliographystyle{aip}
\bibliography{fast_sr.bib}

\begin{thebibliography}{10}

\bibitem{FouMitNeeRaj-RMP-01}
W.~M.~C. Foulkes, L.~Mitas, R.~J. Needs, and G.~Rajagopal,
\newblock Rev. Mod. Phys. {\bf {73}}, 33 (2001).

\bibitem{UmrFil-PRL-05}
C.~J. Umrigar and C.~Filippi,
\newblock Phys. Rev. Lett. {\bf 94}, 150201 (2005).

\bibitem{Sor-PRB-05}
S.~Sorella,
\newblock Phys. Rev. B {\bf 71}, 241103 (2005).

\bibitem{Nightingale:2001:linear_method}
M.~P. Nightingale and V.~Melik-Alaverdian,
\newblock Phys. Rev. Lett. {\bf 87}, 043401 (2001).

\bibitem{TouUmr-JCP-07}
J.~Toulouse and C.~J. Umrigar,
\newblock J. Chem. Phys. {\bf 126}, 084102 (2007).

\bibitem{UmrTouFilSorHen-PRL-07}
C.~J. Umrigar, J.~Toulouse, C.~Filippi, S.~Sorella, and R.~G. Hennig,
\newblock Phys. Rev. Lett. {\bf 98}, 110201 (2007).

\bibitem{TouUmr-JCP-08}
J.~Toulouse and C.~J. Umrigar,
\newblock J. Chem. Phys. {\bf 128}, 174101 (2008).

\bibitem{Sorella:2001:SR}
S.~Sorella,
\newblock Phys. Rev. B {\bf 64}, 024512 (2001).

\bibitem{Sandvik:2007:stoch_opt}
A.~W. Sandvik and G.~Vidal,
\newblock Phys. Rev. Lett. {\bf 99}, 220602 (2007).

\bibitem{Morgan:1990:general_davidson}
R.~B. Morgan,
\newblock J. Comput. Phys. {\bf 89}, 241 (1990).

\bibitem{Linderberg:1977:cholesky}
N.~H.~F. Beebe and J.~Linderberg,
\newblock Int. J. Quantum Chem. {\bf 12}, 683 (1977).

\bibitem{CHANGLANI:2009:cps}
H.~J. Changlani, J.~M. Kinder, C.~J. Umrigar, and G.~K.-L. Chan,
\newblock Phys. Rev. B. {\bf 80}, 245116 (2009).

\bibitem{MEZZACAPO:2009:entangled_plaquettes}
F.~Mezzacapo, N.~Schuch, M.~Boninsegni, and J.~I. Cirac,
\newblock New J. Phys. {\bf 11}, 083026 (2009).

\bibitem{Bajdich:2006:pfaffian}
M.~Bajdich, L.~Mitas, G.~{Drobn\'{y}}, and L.~K. Wagner,
\newblock Phys. Rev. Lett. {\bf 96}, 130201 (2006).

\bibitem{Bajdich:2008:pfaffian}
M.~Bajdich, L.~Mitas, L.~K. Wagner, and K.~E. Schmidt,
\newblock Phys. Rev. B {\bf 77}, 115112 (2008).

\bibitem{Wimmer:2011:pfaffian}
M.~Wimmer,
\newblock arXiv:1102.3440v2  (2011).

\bibitem{Neuscamman:2010:pcps}
E.~Neuscamman, H.~Changlani, J.~Kinder, and G.~K.-L. Chan,
\newblock arXiv:1008.4945v1  (2010).

\bibitem{POPLE:1953:agp}
A.~C. Hurley, J.~Lennard-Jones, and J.~A. Pople,
\newblock Proc. R. Soc. London, Ser. A {\bf 220}, 446 (1953).

\bibitem{COLEMAN:1965:AGP}
A.~J. Coleman,
\newblock J. Math. Phys. {\bf 6}, 1425 (1965).

\bibitem{Poilblanc:1991:tabc}
D.~Poilblanc,
\newblock Phys. Rev. B {\bf 44}, 9562 (1991).

\bibitem{Gros:1996:ibc}
C.~Gros,
\newblock Phys. Rev. B {\bf 53}, 6865 (1996).

\bibitem{ZHANG:2008:hubbard_phase_sep}
C.-C. Chang and S.~Zhang,
\newblock Phys. Rev. B {\bf 78}, 165101 (2008).

\bibitem{Pople:1969:sto-3g}
W.~J. Hehre, R.~F. Stewart, and J.~A. Pople,
\newblock J. Chem. Phys. {\bf 51}, 2657 (1969).

\bibitem{POPLE:1972:6-31g_basis}
W.~J. Hehre, R.~Ditchfield, and J.~A. Pople,
\newblock J. Chem. Phys. {\bf 56}, 2257 (1972).

\bibitem{PIPEK:1989:pipek_mezey_local}
J.~Pipek and P.~G. Mezey,
\newblock J. Chem. Phys. {\bf 90}, 4916 (1989).

\bibitem{Sharma:20xx:sa-dmrg}
S.~Sharma and G.~K.-L. Chan,
\newblock in preparation .

\end{thebibliography}

\end{document}